\documentclass{PoS}

\usepackage{amsmath}
\usepackage{amsfonts}
\usepackage{amssymb}

\title{On distinguishing the direct and spontaneous CP violation in 2HDM }

\ShortTitle{On distinguishing the direct and spontaneous CP violation in 2HDM }

\author{\speaker{Dorota Sokolowska}%
         \thanks{This work is supported in part  by EU Marie Curie Research Training Network HEPTOOLS (contract MRTN-CT-2006-035505) and FLAVIAnet (contract No. MRTN-CT-2006-035482).}\\
        University of Warsaw\\
        E-mail: \email{dsok@fuw.edu.pl}}

\author{Konstantin Kanishev\\
        Sobolev Institute of Mathematics and Novosibirsk State University\\
        E-mail: \email{kanishev@ngs.ru}}

\author{Maria Krawczyk\\
        University of Warsaw\\
        E-mail: \email{krawczyk@fuw.edu.pl}}

\abstract{The most general Two Higgs Doublet Model (2HDM) allows both for the explicit and the spontaneous CP violation in the scalar sector. Here we discuss CP violation in terms of basis-independent quantities and show how using two different sets of known CP-odd weak-basis invariants one can distinguish between CP conservation, explicit CP violation and spontaneous CP violation. The special case of CP violation without CP mixing, in which the neutral Higgs gauge boson interaction respects CP while Higgs self-interaction  violates CP symmetry, is also presented.}

\FullConference{Prospects for Charged Higgs Discovery at Colliders\\
                 16-19 September 2008\\
                 Uppsala, Sweden}

\begin{document}

\section{Introduction}
CP symmetry is one of the most crucial symmetries in particle physics. In the Standard Model CP symmetry is violated through a complex phase in the CKM matrix. However,  this violation is not big enough to lead to the observed baryon asymmetry of the Universe \cite{asymmetry}. Determining all possible sources of CP violation is a fundamental challenge for  high energy physics.

In a CP conserving model both the Lagrangian and the vacuum state conserve CP. There are two possible sources of CP violation - either by parameters of the Lagrangian (direct or explicit violation) or by the vacuum state (spontaneous violation). The 2HDM is one of the simplest extension of the Standard Model that allows both for spontaneous and explicit CP violation \cite{Lee}.

In 2HDM we deal with two doublets of scalar fields with identical quantum numbers, so there is a freedom of choosing the basis in the space of  fields \cite{brancobook,Krawczyk,rebelo,Haber1,Davidson}. Because of this freedom it is convenient  to study CP violation in the 2HDM not in terms of parameters of the potential that depend on the choice of basis but using the CP-odd weak-basis invariant quantities.

We are interested in distinguishing the direct and spontaneous CP violation in 2HDM and for simplicity we focus here on scalar sector only, as in \cite{Lav2}.
We will use the two different sets of weak-basis invariants: $J$-invariants introduced in \cite{Lavoura1} and $I$-invariants \cite{rebelo,Haber1}. They are the analogs of the Jarlskog invariant \cite{Jarlskog} allowing  to distinguish between CP conserving and CP violating models. We show that by combining these invariants one can determine the way  CP is violated - explicitly (directly) or spontaneously.
We discuss separately a special case of CP violation without CP mixing,
%in which  neutral Higgs particles interact with gauge bosons respecting CP, while
where CP violation appears only in the interaction
\cite{Lavoura1}.

\section{The general 2HDM}
Let us consider the most general 2HDM (without the Yukawa interaction)  \cite{Lee,brancobook,Krawczyk,rebelo,Haber1}:
\begin{eqnarray}
 \mathcal{L}& = &\mathcal{L}_{gauge} + \left( D_{\mu} \Phi_{1} \right)^{\dagger} \left( D^{\mu} \Phi_{1} \right) + \left( D_{\mu} \Phi_{2} \right)^{\dagger} \left( D^{\mu} \Phi_{2} \right) - V, \label{Lagrangian} \\
V & = &
\frac{\lambda_{1}}{2} \left(\Phi_{1}^{\dagger} \Phi_{1}\right)^{2} + \frac{\lambda_{2}}{2} \left(\Phi_{2}^{\dagger} \Phi_{2}\right)^{2}  +\lambda_{3} \left(\Phi_{1}^{\dagger} \Phi_{1}\right) \left(\Phi_{2}^{\dagger} \Phi_{2}\right) + \lambda_{4} \left(\Phi_{1}^{\dagger} \Phi_{2}\right) \left(\Phi_{2}^{\dagger} \Phi_{1}\right) \nonumber \\
& & + \left[ \frac{1}{2} \lambda_{5} \left(\Phi_{1}^{\dagger} \Phi_{2}\right) \left(\Phi_{1}^{\dagger} \Phi_{2}\right) + \lambda_{6} \left(\Phi_{1}^{\dagger} \Phi_{1}\right) \left(\Phi_{1}^{\dagger} \Phi_{2}\right) +  \lambda_{7} \left(\Phi_{2}^{\dagger} \Phi_{2}\right) \left(\Phi_{1}^{\dagger} \Phi_{2}\right) + H.c. \right] \nonumber \\
& &  - \frac{1}{2}m_{11}^{2} \left(\Phi_{1}^{\dagger} \Phi_{1}\right) -\frac{1}{2}m_{22}^{2} \left(\Phi_{2}^{\dagger} \Phi_{2}\right) -\frac{1}{2}m_{12}^{2} \left(\Phi_{1}^{\dagger} \Phi_{2}\right) -\frac{1}{2}(m_{12}^{2})^{*} \left(\Phi_{2}^{\dagger} \Phi_{1}\right), \label{V}
\end{eqnarray}
where $\Phi_{1,2}$ are SU(2) doublet with weak hypercharge Y=+1, $\lambda_{1-4}, m_{11}^2, m_{22}^2 \in \mathbb{R}$ and $\lambda_{5-7}, m_{12}^2 \in \mathbb{C}$.

 In the Higgs (Georgi) basis, in which only one doublet acquires the non-zero vacuum expectation value, scalar doublets can be decomposed in the following way:
\begin{equation}
\Phi_1=\left(\begin{array}{c} G^+\\ \frac{ v+\eta_1+i G^0}{\sqrt2}\end{array}\right),
\quad\quad \Phi_2=\left(\begin{array}{c} H^+\\   \frac{\eta_2+i A}{\sqrt2}\end{array}\right). \label{decomposition}
\end{equation}
In this basis we use  special symbols for parameters of the potential, namely   $ \Lambda_i, \mu_{ij}^2$ instead of $\lambda_{i}, m_{ij}^2 $.

There are various CP-transformations possible \cite{brancobook,Nachtman}, which does not change the  kinetic term of (\ref{Lagrangian}). For our purpose it is enough to apply the simplest CP transformation (see also \cite{brancobook}):
\begin{equation}
 \Phi_1 (\vec{x}, t) \to \Phi_1^{\dagger} (-\vec{x}, t), \qquad \Phi_2 (\vec{x}, t) \to \Phi_2^{\dagger} (-\vec{x}, t).
\end{equation}

Under this transformation the neutral fields from the decomposition (\ref{decomposition}) transform as follows: \begin{equation}
\eta_{1,2} \to \eta_{1,2}, \qquad A \to - A,
\end{equation}
so we can identify $\eta_{1}$ and $\eta_2$ as the CP-even fields and $A$ as the CP-odd field
(see also \cite{oneil}).

%\subsection{CP-mixing}
 %In order to discuss the CP phenomena in 2HDM let us have a closer look at the mass matrix and %self-interactions of scalars.
The squared-mass matrix for the neutral fields $(\eta_1, \eta_2, A)$ in the Higgs basis is given by:
\begin{equation}
\mathcal{M}^2= \left(\begin{array}{ccc} M_{11}& M_{12}& M_{13} \\
 M_{12}&M_{22}& M_{23}\\
M_{13} & M_{23} &M_{33}
\end{array}\right)= \left(\begin{array}{ccc} v^2 \Lambda_1& v^2 {\rm Re}\,\Lambda_6& -v^2 {\rm Im}\,\Lambda_6 \\
 v^2 {\rm Re}\,\Lambda_6&\left[v^2 \Lambda_{345}-\mu_{22}^2\right]/2& -v^2 {\rm Im}\,\Lambda_5/2\\
-v^2 {\rm Im}\,\Lambda_6 & -v^2 {\rm Im}\,\Lambda_5/2 & \left[v^2 \tilde\Lambda_{345}-\mu_{22}^2\right]/2
\end{array}\right), \label{massmatrix}
\end{equation}
where $\Lambda_{345} = \Lambda_3 + \Lambda_4 + {\rm Re} \Lambda_5, \,\,\tilde\Lambda_{234} = \Lambda_3 + \Lambda_4 - {\rm Re} \Lambda_5$.
There are two distinct cases here:
\begin{itemize}
\item\emph{CP-mixing.} This is a case with non-zero off-diagonal element $M_{13}$ or $M_{23}$  leading to a mixing between states of different CP properties. Model allows for the CP violation since  physical states $h_i$, being  combinations of $\eta_1, \eta_2, A$, have indefinite CP quantum numbers. This is considered to be a standard way CP is violated in the 2HDM (and in other  models).
\item \emph{no CP-mixing.} When $M_{13} = M_{23} =0$ then there is no CP-mixing and physical (mass-)states have defined CP properties. Neutral Higgs bosons are:  CP-odd  $A$  and CP-even combinations of $\eta_1$ and $\eta_2$ denoted as $h,H$.
    %can be determined in the interaction with gauge bosons. Yet CP violation is possible %due to interactions eg. among Higgs particles.
\end{itemize}
Note that parameter $\Lambda_7$ (in general complex) does not appear in the mass matrix (\ref{massmatrix}) in the Higgs basis, and therefore  it can be related only to the interaction (ie. couplings).

\section{CP-odd weak-basis invariants}
The CP-odd weak-basis invariants are quantities that are invariant under the weak-basis transformation, but change their sign under the CP transformation. In the Standard Model such invariant was introduced by C. Jarlskog for the quark sector \cite{Jarlskog}:
\begin{equation}
J \propto (m_t^2 - m_c^2) (m_t^2 - m_u^2) (m_c^2 - m_u^2) (m_b^2 - m_s^2) (m_b^2 - m_d^2) (m_s^2 - m_d^2) {\rm Im} (V_{ud} V_{cs} V_{us}^{\ast} V_{cd}^{\ast}),  \label{Jarlskog}
\end{equation}
where $m_{i}$ is a mass of $i$-quark, $V_{ij}$ are elements of the CKM matrix.
In the Standard Model $J$ is a only CP-odd invariant and if $J=0$ then CP is conserved. Note, that $J=0$ if two masses of quarks are equal.  We know that in the SM $J \not = 0$  and the CP is violated through the complex phase in the CKM matrix. In the 2HDM the situation is more complicated, however also  here there is a close analog of the Jarlskog invariant, this time for scalars \cite{Lavoura1}:
\begin{equation}
J_1 \propto (m_1^2 - m_2^2) (m_1^2 - m_3^2) (m_2^2 - m_3^2) T_{11} T_{21} T_{31}, \label{J1mass}
\end{equation}
where $m_i, i=1,2,3$ are masses of neutral scalars and  $T_{ij}$ -- elements of rotation matrix between the mass-states and  $(\eta_1, \eta_2, A)$ fields. Also here if two masses of neutral Higgs bosons are equal $J_1=0$. It is important to point out that in the 2HDM $J_1 = 0$ does not imply CP conservation in the scalar sector, in contrast to the Standard Model case. Vanishing of the $J_1$ is now a necessary but not sufficient condition for CP conservation \cite{Lavoura1}.
%The clue is that  here there are more relevant CP-odd invariants.

\subsection{$J$-invariants}
The $J_1$ invariant (\ref{J1mass}) is the only CP-odd invariant in 2HDM which can be constructed from the squared-mass matrix \cite{Lavoura1}. It can be written also in two other forms by using elements of squared-mass matrix (\ref{massmatrix}) before diagonalization and by parameters of the potential $V$ (\ref{V}), respectively:

\begin{equation}
J_1 = M_{12} M_{13} (M_{22}-M_{33}) + M_{23} \left( M_{13}^2 - M_{12}^2 \right) = - 8 v^6 {\rm Im}\,(\Lambda_5^{\ast} \Lambda_6^2). \label{J1}
\end{equation}

 From the above form of $J_1$ it is easy to see that if any two of three off-diagonal elements of the matrix $\mathcal{M}^2$ vanish then $J_1 = 0$.  However, even when $J_1=0$ CP violation is possible \cite{Lavoura1}, signalizing by two additional  CP-odd invariants containing   $\Lambda_7$, the only complex parameter which is absent in $\mathcal{M}^2$. So, the set of relevant CP-odd weak-basis invariant is  \footnote{Only two of three $J$-invariants are independent in the general 2HDM without fermions \cite{Lavoura1}.}
%Two distinct cases  with $J_1$ are:
 %in two different ways with and without mixing between states with different CP properties
%It happens that this parameter enters  ,
%which govern the interaction between Higgs particles []:
\begin{equation}
J_1=- 8 v^6 {\rm Im}\,(\Lambda_5^{\ast} \Lambda_6^2), \qquad J_2 = - 4 v^4 {\rm Im}\,(\Lambda_5^{\ast} \Lambda_7^2), \qquad J_3 = 2 \sqrt{2} v^3 {\rm Im}\,(\Lambda_7^{\ast} \Lambda_6).
 \end{equation}

To have CP conserving model all $J$-invariants must vanish \cite{Lavoura1}. If any of those three invariants does not vanish, then there is CP violation in the model. As we already discussed there are two cases with and without CP-mixing, which in terms of $J_{1,2,3}$ can be described as follows:
\begin{itemize}
\item If $J_1 \not = 0$ ($M_{13}, M_{23} \not = 0$) then there is mixing between states of different CP properties and physical states $h_1, h_2, h_3$ have no defined CP quantum numbers.
\item If $J_1 = 0$ and $J_{2,3} \not = 0$ we have CP violation from interactions even if there is no mixing between states. This case will be discussed further in section 4.
\end{itemize}
 It is worth noticing that all that these findings, in particular the fact of possible CP violation without CP-mixing, are valid not only in the Higgs basis \cite{Lavoura1}.
 Note, however that CP violation without CP mixing  can not be realized in the 2HDM with soft violation of $Z_2$ symmetry.

 It is important to realize that $J$-invariants do not distinguish between explicit and spontaneous CP violation. So, in order to pin down   kind of CP violation  other types of invariants are needed.
\subsection{$I$-invariants}
The potential (\ref{V}) can be written in the following form \cite{brancobook, Davidson}:
\begin{equation}
V = Y_{a\bar{b}} \Phi_{\bar{a}}^{\dagger} \Phi_{b} + \frac{1}{2} Z_{a \bar{b} c \bar{d}} (\Phi_{\bar{a}}^{\dagger} \Phi_b ) (\Phi_{\bar{c}}^{\dagger} \Phi_d ). \label{Vbranco}
\end{equation}
Four  CP-odd weak-basis invariants can be build from parameters $Y_{a\bar{b}}, Z_{a \bar{b} c \bar{d}} $ of $V$ (\ref{Vbranco}) \cite{rebelo,Haber1}:

\begin{equation}
\begin{array}{rclcrcl}
I_{1} &=& \mathrm{ Im} (Z_{a\bar{c}}^{(1)} Z_{e \bar{b}}^{(1)} Z_{b \bar{e} c \bar{d}} Y_{d \bar{a}}), &&  I_{2} &=& \mathrm{Im} (Y_{a \bar{b}} Y_{c \bar{d}} Z_{b \bar{a} d \bar{f}} Z_{f \bar{c}}^{(1)}),\\
I_{3} &=& \mathrm{Im} (Z_{a\bar{c} b \bar{d}} Z_{b \bar{f}}^{(1)} Z_{d \bar{h}}^{(1)} Z_{f\bar{a} j \bar{k}} Z_{k\bar{j} m \bar{n}} Z_{n\bar{m} h \bar{c}}), && I_4 &=& \mathrm{Im}(Z_{a\bar{c} b \bar{d}} Z_{c\bar{e} d \bar{g}} Z_{e\bar{h} f \bar{q}} Y_{g \bar{a}} Y_{h \bar{b}} Y_{q \bar{f}}).
% \scriptstyle{I_1}&=& \scriptstyle{2 (|\lambda_6|^2 - |\lambda_7|^2) {\rm Im}[Y_{12} (\lambda_6^{\ast} + \lambda_7^{\ast})] + (\lambda_1 - \lambda_2) [{\rm Im}(Y_{12} \Lambda^{\ast}) - {\rm Im}[Y_{12} \lambda_5^{\ast}(\lambda_6 + \lambda_7)]]} & \\
% & \scriptstyle{ + (Y_{11} - Y_{22}){\rm Im}[\lambda_5^{\ast} (\lambda_6 + \lambda_7)^2] - (\lambda_1 - \lambda_2) [{\rm Im}(\lambda_7^{\ast} \lambda_6 ]}&
 \end{array}
\end{equation}
(Here $Z^{(1)}_{a\bar b}$ denotes a combination of the parameters $Z_{a\bar b c \bar d}$.)
Note that in $I_i$ both quartic and quadratic parameters enter in contrast to $J_i$,  derived for physical states, which can be expressed  by quartic parameters $(\Lambda_i)$ only.

%The derivation of $I_i$ and relevant theorems was done in \cite{Haber1}.
%, so let us just summarize the properties of $I$-invariants.
 The potential is explicitly CP conserving if and only if all $I_i$ vanish \cite{rebelo,Haber1}. This means that there exists ,,a real basis'' of fields $\Phi_1, \Phi_2$ in which all $\lambda_i, m_{ij}^2$ are real. In such case still CP can be violated spontaneously, since
 %However, as stated before, this does not mean that the model is CP conserving.
 $I$-invariants are not sensitive to vacuum expectation value of fields. So, if $\forall I_i = 0$ then CP symmetry in the model can be either conserved (both explicitly and spontaneously) or violated spontaneously by the vacuum state.

\subsection{Distinguishing between various kinds of CP violation}

As we already discussed $J$- and $I$-invariants are sensitive to the different aspects of CP violation in the scalar sector of 2HDM. In particular we see that they have different sensitivity  for the spontaneous CP violation ($I_i$ have none). Combining the information provided by $J$- and $I$-invariants allows us to distinguish between conservation and violation of CP symmetry and if CP is violated to establish the pattern of this
 violation. A comparison of $J$- and $I$-invariants is shown in the table \ref{table}.

% \begin{itemize}
%  \item If $\forall J_{i} = 0 $ then we have CP conserving case. In this case all $I_i$ must be also 0.
% \item If $\exists J_{i} \not = 0 $ then we have CP violation and $I_i$ allow us to distinguish between explicit and spontaneous violation.
% \begin{itemize}
% \item If $\forall I_{i} = 0 $ then we have spontaneous CP violation and there exists a basis in which all the parameters are real. The CP symmetry is violated by the vacuum state.
% \item If $\exists I_{i} \not = 0 $ then we have explicit CP violation and in every basis some of the parameters are complex.
% \end{itemize}
% \end{itemize}

\begin{table}[h]
\caption{A comparison of $J$- and $I$-invariants in the 2HDM} \label{table}
\begin{center}
\begin{tabular}{ || c | c | c ||} \hline
CP properties of 2HDM & J-invariants & I-invariants\\[0.3cm] \hline \hline
CP explicitly violated & $\exists J_{i} \not = 0 $ & $\exists I_{i} \not = 0 $\\[0.2cm] \hline
CP spontaneously violated &$\exists J_{i} \not = 0 $ & $\forall I_{i} = 0 $ \\[0.2cm] \hline
CP conserved & $\forall J_{i} = 0 $ & $\forall I_{i} = 0 $ \\ \hline
\end{tabular}
\end{center}
\end{table}

\vspace{-1cm}
\section{CP violation without CP mixing}

Let us now consider a special case, which we mentioned earlier, when physical states with defined CP properties have CP-violating interaction. In such model (in the Higgs basis) let us take
%$${\rm Im}\, \Lambda_5 = {\rm Im}\, \Lambda_6 = 0.$$
\[{\rm Im}\, \Lambda_5 = {\rm Im}\, \Lambda_6 = 0.\]
 %and the neutral squared-mass matrix in the basis $(\eta_1, \eta_2, A)$ is given by:
%\begin{equation}
%\mathcal{M}^2 = \left(\begin{array}{ccc} v^2 \Lambda_1& v^2 {\rm Re}\,\Lambda_6& 0 \\
% v^2 {\rm Re}\,\Lambda_6&\left[v^2 \Lambda_{345}-\mu_{22}^2\right]/2& 0\\
%0 & 0 &\left[v^2 \tilde\Lambda_{345}-\mu_{22}^2\right]/2
%\end{array}\right).
%\end{equation}
Due to the chosen values of $\Lambda_5$ and $\Lambda_6$ there is no mixing between states of different CP and the $J_1$ invariant (\ref{J1}) is equal to zero. After standard diagonalization we get states $h$, $H$ and $A$ with defined CP properties.

The interaction of these particles $h,H,A$ with gauge bosons is described by
\begin{equation}
\mathcal{L}_{s-g} \supset \frac{g^2 v}{2} \chi_{h}^{V} h W^{+} W^{-}  + \frac{g^2 v}{2} \chi_{H}^{V} H W^{+} W^{-} + \frac{g^2 v}{2} \chi_{A}^{V} A W^{+} W^{-},
\end{equation}
\begin{equation}
 \chi_{h}^{V} = \cos \alpha, \qquad \chi_{H}^{V} = \sin \alpha , \qquad \chi_{A}^{V} = 0.
\end{equation}
Here $\alpha$ is a mixing angle between $\eta_1$ and $\eta_2$ fields. The $\chi_i$ are the relative couplings with respect to the Standard Model coupling between the SM Higgs boson and $W/Z$, with a sum rule $(\chi_{h}^{V} )^2 + (\chi_{H}^{V} )^2 + (\chi_{A}^{V} )^2 =~1$. All these couplings  are like in the CP conserving case, in contrast to the case of CP mixing, where
%(Note, that if physical states do not have defined CP properties
all couplings $\chi_{i}^V $ for $h_i$ are in principle nonvanishing, in particular
 $\chi_{3}^{V} \not = 0$.

Although the  mass-squared matrix and interactions with gauge bosons point to the CP conservation,  the remaining nonvanishing for ${\rm Im}\, \Lambda_7 \not = 0$  $J$-invariants
\begin{equation}
{\rm Im}\, \Lambda_7 \not = 0 \Rightarrow J_2, J_3 \not = 0,
\end{equation}
%Therefore, we conclude that although $J_1 = 0$,
ensure us that there is the CP violation in the considered (no CP-mixing) case.
 This CP violation shows up for example in the trilinear self-interaction of physical scalars   $\mathcal{L}_{self}$:
\begin{eqnarray}
\mathcal{L}_{self} & \supset & -\frac{1}{2} {\rm{Im}} \Lambda_7 v \, A A A - \frac{1}{2} {\rm{Im}} \Lambda_7 v \sin^2 \alpha  \, A h h - \frac{1}{2} {\rm{Im}} \Lambda_7 v \cos^2 \alpha \, A H H ,  \nonumber \\
&& + {\rm{Im}} \Lambda_7 v \cos \alpha \sin \alpha \, A h H -{\rm{Im}} \Lambda_7 v  \, A H^{+} H^{-}.
\end{eqnarray}

We see that if ${\rm{Im}} \Lambda_7 \neq 0$ there are possible  couplings with odd number of the CP-odd field $A$, which cannot occur in the CP conserving 2HDM. For example,  new decay channels appear for $A$: $A \to hh, \,\, HH, \,\,H^+ H^-$. $J$-invariants tell us about CP violation in the model, however if we want to know what  kind of CP violation occurs here we need to use $I$-invariants.
% $I$-invariants will give us the answer what is the source of CP violation.

%\[A A A \propto -v {\rm{Im}} \Lambda_7, \qquad A h h \propto -{\rm{Im}} \Lambda_7 v \sin^2 \alpha\]
%\[A H H \propto -{\rm{Im}} \Lambda_7 v \cos^2 \alpha, \qquad A h H \propto {\rm{Im}} \Lambda_7 v \cos \alpha \sin \alpha \]
%\[A H^{+} H^{-} \propto  -{\rm{Im}} \Lambda_7 v  \]
% \[ {\color{blue} A A A \propto}  {\color{red}-{\rm{Im}} \Lambda_7} {\color{blue}v}, \qquad {\color{blue} A h h \propto} {\color{red}-{\rm{Im}} \Lambda_7} {\color{blue}v \sin^2 \alpha,} \]
% \[ {\color{blue} A H H \propto} {\color{red}-{\rm{Im}} \Lambda_7} {\color{blue}v \cos^2 \alpha,} \qquad {\color{blue} A h H \propto} {\color{red} {\rm{Im}} \Lambda_7} {\color{blue} v \cos \alpha \sin \alpha,} \]
% \[ {\color{blue} A H^{+} H^{-} \propto} {\color{red} -{\rm{Im}} \Lambda_7} {\color{blue}v}  \]

\section{Summary}

We studied the CP violation in the 2HDM with the aim to distinguish between the explicit and spontaneous form of this violation. We used two sets of well known CP-odd weak-basis invariants and we found that both of them are needed to pin down the nature of CP violation.

We discuss  a special case of CP violation without CP mixing, which is in contradiction with the usual treatment, where CP violation in the 2HDM is considered as being equivalent to the mixing between states with different CP properties. We found a case in which there is no CP mixing and the interaction of neutral Higgs bosons with gauge bosons preserves CP, however the Higgs self-interaction  violates CP symmetry. This results in non-zero vertices with odd number of A (eg.~$A \to H^+ H^-$).

\paragraph{Acknowledgments}
We are grateful for discussion with H. Haber and I. Ivanov.

\end{document}